\documentclass[aps,prb,twocolumn]{revtex4}

\usepackage{amsmath}
\usepackage{color}
\usepackage{amssymb}

\usepackage{graphicx}

\begin{document}

\title{Whether there is the intrinsic Hall effect in a multi-band superconductor?}

\author{V.P.~Mineev}
\affiliation{Service de Physique Statistique, Magn\'{e}tisme et
Supraconductivit\'{e}, Institut Nanosciences et Cryog\'{e}nie, UMR-E CEA/UJF-Grenoble1, F-38054 Grenoble, France}

\begin{abstract}
The interplay of the microscopic models and the symmetry considerations in application to the superconducting state of putative chiral superconductor Sr$_2$RuO$_4$ are presented.  
There is demonstrated that chiral  two-band superconducting ordering recently proposed as  candidate for the
superconducting state in this material and having the long-expected property such as  the intrinsic Hall effect
does not exist as energetically disadvantageous  in comparison with the states not supporting the Hall effect.
Other chiral  multi-band superconducting states with intraband pairing do not   support an intrinsic Hall properties as well.
The superconducting states with direct interband pairing could serve as the probable candidates for the Hall effect existence.

KEYWORDS: Sr$_2$RuO$_4$, chiral superconductivity

\end{abstract}

\date{\today}
\maketitle
\section{Introduction}
During about two decades the layered perovskite material Sr$_2$RuO$_4$  attracts a lot of attention of low temperature community (for the recent review see \cite{Maeno}). This is mostly due to its putative chiral p-wave superconductivity which is still controversial but supported by several significant experimental observations \cite{Kallin}. The chiral superconductivity breaks the time reversal symmetry. Hence, as a ferromagnet, it has to reveal the anomalous Hall effect. Indeed, there was 
measured the polar Kerr effect related to the nonzero value of the Hall conductivity at infrared frequencies \cite{Xia}.  
On the other hand there was clearly demonstrated 
that the Hall conductivity in a translationally invariant chiral superconductor  is equal to zero \cite{Read}.
  In view of this general statement
  the proposed explanations of the Hall  effect  have been purely {\it extrinsic} arising from impurity scattering\cite{Lutchin,Goryo}.

Among the attempts to point out the microscopic mechanism of chiral p-wave superconductivity  there is recently published model\cite{Raghu} possessing specific properties and based on  quasi-one-dimensional character  of 
two  conducting bands in this material.  (see also \cite{Takimoto}).  This model has revealed the important consequence, namely - the {\it intrinsic} Hall effect arising from interband transitions involving time-reversal symmetry breaking chiral Cooper pairs as it was analytically demonstrated in the paper 
\cite{Taylor}. 
In parallel with the paper \cite{Taylor} there were published the results of numerical calculations of the intrinsic Hall effect in a multi-band   
superconducting state breaking the time reversal symmetry \cite{Wysokinski}.

A superconducting state realizing in a compound possesses  definite properties dictated by the host crystal symmetry\cite{Volovik,Mineev}. 
And a particular microscopic model for superconducting state  built theoretically 
should be constructed in correspondence with the symmetry requirements.  Here we discuss the  properties
of two-band superconducting state introduced in the papers \cite{Raghu, Taylor}. More generally, we discuss the properties of one component and multi-component superconducting states in a two-band tetragonal superconductor.
There is demonstrated that not a single state  with intraband pairing obeys the properties necessary to support  the  intrinsic Hall effect. On the other hand  
the Hall effect in the superconductors with direct interband pairing still deserves the further studies.

The work \cite{Taylor}  presents the explicit analytic expression for the Hall conductivity in terms of exactly formulated microscopic model taking  into account not only inter-band Cooper pairs transitions  
but  formally also the direct inter-band pairing. To formulate the conditions for existence of the intrinsic Hall effect
we begin with rewriting several equations from the paper \cite{Taylor}.
Then we write the Landau free energy expansions and consider the symmetry properties  for the one-component and two-component superconducting states
in two band  superconductor. The results are formulated in the Conclusion.

\section{Hall conductivity}
We shall discuss unitary equal spin pairing superconducting states.
This  case  all the calculations for the spin up-up $|\uparrow\uparrow\rangle$  and spin down-down $|\downarrow\downarrow\rangle$ order parameter components  are separated and equivalent each other.
There are also  another type of  superconducting states having only one  spin component $|\uparrow\downarrow\rangle+|\downarrow\uparrow\rangle$ corresponding to the equal spin pairing with spins perpendicular to the spin quantization axis usually  chosen parallel to the tetragonal axis $\hat z$. This case all the written equation are related to this single component.

The intrinsic Hall conductivity
\begin{equation}
\sigma_H(\omega)=-\frac{1}{2i\omega}\lim_{{\bf q}\to 0}[\pi_{xy}({\bf q},\omega)-\pi_{yx}({\bf q},\omega)]
\end{equation}
 was found in the paper\cite{Taylor} as analytic continuation of antisymmetric part of the Matsubara current correlator 
\begin{equation}
\pi_{xy}({\bf q},\nu_m)=e^2T\sum_{{\bf k},\omega_n}Tr[{\bf  v}_x{\bf G}({\bf k},\omega_n){\bf  v}_y{\bf G}({\bf k}+{\bf q},\omega_n+\nu_m)]
\end{equation}
to the real frequencies $i\nu_m\to\omega+i0^+$.
The matrix of electron velocities is determined as derivative of dispersion of two-band 
noninteracting Bloch electrons
\begin{equation}
{\bf  v}=\left (\begin{array}{cc}{\bf v}_1({\bf k})\hat\tau_0 &{\bf v}_{12}({\bf k})\hat\tau_0 \\  {\bf v}_{12}({\bf k}) \hat\tau_0
&  {\bf v}_{2}({\bf k})\hat\tau_0\end{array}\right )=\frac{\partial}{\partial {\bf k}}\left (\begin{array}{cc}\xi_1({\bf k})\hat\tau_0 &  \varepsilon_{12}({\bf k})\hat\tau_0 \\  \varepsilon_{12}({\bf k}) \hat\tau_0
&  \xi_{2}({\bf k})\hat\tau_0\end{array}\right ).
\end{equation}
Here,
\begin{eqnarray}
&\xi_{1}+\mu=-2t\cos k_x-2t^\perp\cos k_y,\nonumber\\
&\xi_{2}+\mu=-2t\cos k_y-2t^\perp\cos k_x
\end{eqnarray}
are the dispersions for the Bloch states in Sr$_2$RuO$_4$ forming bands (numbered by indices 1,2) constructed from the $xz$ and $yz$ ruthenium orbitals correspondingly. To avoid misunderstanding we note, that these bands are often called in the paper \cite{Taylor} by orbitals unlike to
real $\alpha$ and $\beta$ bands in this compound those dispersion laws are obtained after usual diagonalization procedure  taking into account the interband coupling 
\begin{equation}
 \varepsilon_{12}=-2t^{\prime\prime}\sin k_x\sin k_y.
\end{equation}

The 4x4 Green functions for two-band superconductor with equal spin pairing for each spin projection is  defined through its inverse matrix
\begin{eqnarray}
{\bf G}^{-1}({\bf k},\omega_n)=
\left (\begin{array}{cc} i\omega_n\hat\tau_0-\xi_1\hat\tau_3-\hat\Delta_{1}
& -\varepsilon_{12}\hat\tau_3+\hat\Delta_{12}\\
 -\varepsilon_{12}\hat\tau_3+\hat\Delta_{12}&i\omega_n\hat\tau_0-\xi_2\hat\tau_3-
\hat \Delta_2
 \end{array}\right ).
 \label{Green}
\end{eqnarray}
Here,
$\hat\tau_0,~\tau_3$ are the Pauli matrices 
 in the particle-hole space, and the order parameter  matrices
for intraband   and direct  interband  pairing are correspondingly
\begin{equation}
\hat\Delta_{1,2}=\left (\begin{array}{cc}0&\Delta_{1,2}({\bf k})\\
 \Delta_{1,2}^\star({\bf k})&0\end{array}\right ),\hat\Delta_{12}=\left (\begin{array}{cc}0&\Delta_{12}({\bf k})\\
 \Delta_{12}^\star({\bf k})&0\end{array}\right ).
\end{equation}
Following the paper \cite{Taylor} we we did not include the interband spin-orbital interaction.

The calculation performed in the paper\cite{Taylor} yields 
\begin{widetext}
\begin{equation}
\pi_{xy}({\bf q}=0,\nu_m)=
4e^2T\nu_m\sum_{{\bf k},\omega_n}\frac{(2\omega_n+\nu_m)^2[({\bf v}_2-{\bf v}_1)\times{\bf v}_{12}]_z\left [\varepsilon_{12}
\mbox{Im}(\Delta_1^\star\Delta_2)+\xi_1\mbox{Im}(\Delta_2^\star\Delta_{12})-\xi_2\mbox{Im}(\Delta_1^\star\Delta_{12})\right ]}
{(\omega_n^2+E_+^2)(\omega_n^2+E_-^2)[(\omega_n+\nu_m)^2+E_+^2][(\omega_n+\nu_m)^2+E_-^2]}.
\label{pi}
\end{equation}
\end{widetext}
Here $E_{\pm}$ are solutions of the equation  
det${\bf G}^{-1}({\bf k},\omega_n)=(\omega_n^2+E_+^2)(\omega_n^2+E_-^2)$. Unlike to Eqs.(7) and (8) in the paper\cite{Taylor} the above expression is written as it is before the summation over Matsubara frequencies.

Formula (\ref{pi}) contains the terms originating from intraband pairing multiplied by the interband matrix element  $\varepsilon_{12}$ and the terms proportional to the direct interband pairing order parameter $\Delta_{12}$. The later are usually ignored in the theory of two band superconductivity due the mismatching of oppositely directed momenta of the Cooper pairs formed by the electrons from different bands.\cite{Suhl}
We discuss, first,  the terms containing the interband transition matrix element. Then, we have
\begin{eqnarray}
&[({\bf v}_2-{\bf v}_1)\times{\bf v}_{12}]_z\varepsilon_{12}
\mbox{Im}(\Delta_1^\star\Delta_2)=\nonumber\\
&-8(t-t^\perp)(t^{\prime\prime})^2(\sin^2k_x\cos k_y+\sin^2k_y\cos k_x)\times\nonumber\\
&\sin k_x\sin k_y\mbox{Im}(\Delta_1^\star\Delta_2)
\label{angledep}
\end{eqnarray}

The integration in Eq.(\ref{pi}) is produced over the Brillouin cell $\sum_{\bf k}=\int_{-\pi}^\pi \int_{-\pi}^\pi dk_xdk_y$.
The denominator in this formula depending from $E_+$ and $E_-$ is even function both in respect  to $k_x$ and $k_y$. Hence, whether there is   nonzero value 
of $\pi_{xy}({\bf q}=0,\nu_m)$, it   is determined by the product 
\begin{equation}
\sin k_x\sin k_y\mbox{Im}(\Delta_1^\star\Delta_2)
\label{product} 
\end{equation}
in the last line of Eq.(\ref{angledep}).
Now, one must fixe the two-band superconducting state for which the integral over Brillouin zone  is not zero. 

\section {Symmetry considerations} 
\subsection{1D representations}
The authors of paper\cite{Raghu}
 discuss the superconducting state making use the  model of two equivalent bands and corresponding pairing interaction with orthorhombic symmetry  ignoring the interband Cooper pairs scattering \cite{Suhl} $\sum_{{\bf k}{\bf k}'}V_{12}({\bf k},{\bf k}')
 a_{1-{\bf k}}^\dagger a_{1{\bf k}}^\dagger a_{2{\bf k}'}a_{2-{\bf k}'}$ but taking into account the interband coupling $ \varepsilon_{12}$. The orthorhombic point group has only one dimensional representations.
 The superconducting states in two bands\cite{Raghu} are transformed according different 1-D representations of the orthorhombic group, namely, as $\Delta_1=\eta_1\sin k_x=|\eta_1|e^{i\varphi_1}\sin k_x$ in the first band and as $\Delta_2=\eta_2\sin k_y=|\eta_2|e^{i\varphi_2}\sin k_y$ in the second one. This case, the Landau free energy expansion has the following form
 \begin{eqnarray}
 &F=\alpha(|\eta_1|^2 +|\eta_2|^2)+\beta(|\eta_1|^4 +|\eta_2|^4)+\beta_{12}|\eta_1|^2 |\eta_2|^2\nonumber\\
 &+\tilde\beta_{12}\left [\eta_1^2(\eta_2^\star)^2+(\eta_1^\star)^2\eta_2^2\right ].
\label{GL1}
 \end{eqnarray}
 The coefficients $\alpha=\alpha_0(T-T_{c0})$ and, hence, the critical temperatures 
 for both bands are equal due the band equivalence. Now, we see, that if the coefficient  $\beta_2$ is positive (this is  indeed the case in the model\cite{Raghu}), then the phase shift between the  superconducting order parameters in bands 1 and 2 is $\varphi_1-\varphi_2=\pi/2$.
 The product (\ref{product}) proves to be equal to $|\eta_1||\eta_2|\sin^2 k_x\sin^2 k_y$ with obviously nonzero integral over the Brillouin zone.
 
In fact, it is incorrect to consider the superconducting states in two different bands of tetragonal crystal as relating to the one-dimensional representations of  orthorhombic group. Such consideration in  terms of a microscopic model means that the Hamiltonian is chosen as not obeying the necessary symmetry  conditions. A superconducting state in tetragonal crystal should be transformed according to some representation of tetragonal point group.

The construction, from that follows the  coincidence of critical temperatures for the superconducting states transforming according to different representations, is unrealistic. The critical temperatures for the phase transition to the superconducting states relating to different representation are inevitably different. From the point of view of microscopic theory it follows from the difference of bands in the normal state, in given case, the difference between $\alpha$ and $\beta$ bands in strontium ruthenate\cite{Mackenzie}. The model \cite{Raghu}  is relevant in  the unphysical situation when the superconducting critical temperature is larger than the band hybridization energy $T_c>>\varepsilon_{12}$. Then,  at critical temperature, the bands can be treated as equivalent. In reality  the opposite inequality takes place.  Hence,  the $xz$ and $yz$ orbitals  (bands 1 and 2) should be taken as hybridized and forming different $\alpha$ and $\beta$ bands even at temperatures much larger than the temperature of the superconducting transition. 
We, however, continue our discussion  because  even at proper treatment of superconducting state realizing in presence of  the interband Cooper pairs scattering \cite{Suhl} $\sum_{{\bf k}{\bf k}'}V_{12}({\bf k},{\bf k}')
 a_{1-{\bf k}}^\dagger a_{1{\bf k}}^\dagger a_{2{\bf k}'}a_{2-{\bf k}'}$ the  Eqs. (\ref{pi}) and (\ref{angledep}) are still valid.

According to general rules of nonconventional superconductivity theory the superconducting states arising in a two band metal at the same critical temperature  must be related to the same superconducting class\cite{Volovik}. The Landau functional for one-component states has the following form
\begin{eqnarray}
 &F=\alpha_1|\eta_1|^2 +\alpha_2|\eta_2|^2+\gamma\left [\eta_1\eta_2^\star+\eta_1^\star\eta_2\right ]\nonumber\\
&+\beta_1|\eta_1|^4 +\beta_2|\eta_2|^4+\beta_{12}|\eta_1|^2 |\eta_2|^2\nonumber\\
 &+\tilde\beta_{12}\left [\eta_1^2(\eta_2^\star)^2+(\eta_1^\star)^2\eta_2^2\right ].
 \label{L2}
 \end{eqnarray}
 The coefficients $\alpha_1=\alpha_{10}(T-T_{c1})$, $\alpha_2=\alpha_{20}(T-T_{c2})$,  $T_{c1}$ and $T_{c2}$  are the critical temperatures in the absence of interband pair scattering \cite{Suhl}. 
The term $\gamma\left [\eta_1\eta_2^\star+\eta_1^\star\eta_2\right ]$ in Eq. (\ref{GL1})  is  absent simply due the fact that the superconducting states in different bands belong to different representations.  In terms of microscopic theory it means that the interband Cooper pairs transitions are present but only starting from the forth order terms in respect to the order parameter products.  These transitions produce  in 
Eqs. (\ref{GL1}), (\ref{L2})  the band-mixing terms $\beta_{12}|\eta_1|^2 |\eta_2|^2$ and $\tilde\beta_{12}\left [\eta_1^2(\eta_2^\star)^2+(\eta_1^\star)^2\eta_2^2\right ]$. The microscopic calculation shows that the fourth and the higher order  mixing terms are  proportional to the powers of $\varepsilon_{12}$ and completely absent 
at $\varepsilon_{12}=0$.
 On the contrary,  for the superconducting states belonging to the same superconducting class the second order, that is $\gamma\left [\eta_1\eta_2^\star+\eta_1^\star\eta_2\right ]$ term, always exists independently of presence or absence of hybridization $\varepsilon_{12}$. Note, that in presence  $\gamma\left [\eta_1\eta_2^\star+\eta_1^\star\eta_2\right ]$ term,
 the phase shift between the bands order parameters proves to be $0$   for $\gamma<0$ or  $\pm\pi$ for $\gamma>0$.

 The symmetry statement that superconducting states in different bands are obligatory related to the same representation has the simple energetic explanation.
Indeed, even at $\alpha_1=\alpha_2$ a superconducting state with $\gamma$ term has higher critical temperature $T_c=T_{c0}+\frac{|\gamma|}{\alpha_0}$ of the phase transition to the superconducting state than  the critical temperature $T_{c0}$ in the absence $\gamma$ term.  
 So, the band order parameters  belong to the same representation.

 The order parameters  for the superconducting  states in a metal with tetragonal symmetry are listed in  Tables 3 and 4  in the book \cite{Mineev}. To take into account the translational symmetry one must simply substitute $k_{x(y)}$ by $\sin k_{x(y)} $ in these Tables.  It is easy to check\cite{footnote} that for a pair of
superconducting states transforming according to the same one-dimensional representation of the tetragonal group
the integral over $k_x,k_y$ from expression (\ref{product}) vanishes identically yielding
$\pi_{xy}({\bf q}=0,\nu_m)=0$. Hence, all these states do not support the intrinsic Hall effect. 
\subsection{2D representation}
Now one needs to investigate the superconducting state transforming according to two-dimensional representation. The band order parameters for these states 
have the form
\begin{eqnarray}
&\Delta_{1}=\eta_{1x}\psi_{1x}+\eta_{1y}\psi_{1y},\nonumber\\
&\Delta_{2}=\eta_{2x}\psi_{2x}+\eta_{2y}\psi_{2y}
\end {eqnarray}
where the order parameter  amplitudes are vectors  $\vec\eta_1=(\eta_{1x}$, $\eta_{1y})$, $\vec\eta_2=(\eta_{2x},\eta_{2y})$. The functions $\psi_{1(2)x}$ and $\psi_{1(2)y}$ of $E_{u}$ irreducible representation are transformed as $\sin k_x$ and $\sin k_y$ correspondigly.
The Ginzburg-Landau free energy for the two-band superconducting state transforming according to two-dimensional representation has the following form
\begin{eqnarray}
&F=\alpha_1{\vec\eta}_1{\vec \eta}_1^\star +\beta_1({\vec\eta}_1{\vec \eta}_1^\star)^2+\tilde\beta_1|{\vec\eta}_1{\vec \eta}_1|^2+
\beta_1^\prime(|\eta_{1x}|^4+|\eta_{1y}|^4)\nonumber\\
&+\alpha_2{\vec\eta}_2{\vec \eta}_2^\star +\beta_2({\vec\eta}_2{\vec \eta}_2^\star)^2+\tilde\beta_2|{\vec\eta}_2{\vec \eta}_2|^2+
\beta_2^\prime(|\eta_{2x}|^4+|\eta_{2y}|^4)\nonumber\\
&+\gamma({\vec\eta}_1{\vec \eta}_2^\star+{\vec\eta}_1^\star{\vec \eta}_2)\nonumber\\
&+\beta_3({\vec\eta}_1{\vec \eta}_1^\star)({\vec\eta}_2{\vec \eta}_2^\star)+\beta_4[({\vec\eta}_1{\vec \eta}_1)({\vec\eta}_2^\star{\vec \eta}_2^\star)+({\vec\eta}_1^\star{\vec \eta}_1^\star)({\vec\eta}_2{\vec \eta}_2)]\nonumber\\
&+\beta_5|{\vec \eta}_1{\vec \eta}_2^\star|^2+\beta_6({\vec\eta}_1{\vec \eta}_1^\star+{\vec\eta}_2{\vec \eta}_2^\star )({\vec\eta}_1{\vec \eta}_2^\star+{\vec\eta}_1^\star{\vec \eta}_2)
\label{GL}
\end{eqnarray}
Here, $\alpha_1=\alpha_{10}(T-T_{c1}),~\alpha_2=\alpha_{20}(T-T_{c2})$,  $T_{c1}$ and $T_{c2}$ are the critical temperatures in the absence of interband pair scattering. 
Again, the fourth  order band mixing terms have nonzero values only in presence of the band hybridization term $\varepsilon_{12}$ in Eq. (\ref{Green}).
We remind that  for each spin component of an equal pairing state the free energy expansion has the same form.

The Landau  free energy ({\ref{GL}) has completely general character determined only by the crystal symmetry and the dimensionality of representation. The coefficients $\alpha_1,\dots \beta_1,\dots\gamma$ can be determined in frame  of some microscopic model of pairing. 
However, even without knowledge  of particular values of the coefficients one can make several important conclusions concerning the properties of superconducting state \cite{Volovik}. For example, if we discuss just  one-band superconductivity,
where the corresponding Landau free energy is given by the first line in Eq.(\ref{GL}),  one can conclude that all the superconducting states with 
$\vec\eta_1\propto(1,0), (1,1),(1,i),\dots$ have the same critical temperature. The choice between them can be made with aid of the fourth order terms. For example \cite{Volovik}, for coefficients $\tilde\beta_1>0$ and $\beta_1^\prime>-2\tilde\beta_1$ the most profitable superconducting state proves to be  time reversal symmetry breaking state $\vec\eta_1\propto(1,\pm i)$.   The beta-coefficients found in frame of weak coupling theory with arbitrary shape of the Fermi surface ($\beta_1= 2\tilde\beta_1>0, \beta_1^\prime=0$)     support this conclusion (see for instance \cite{Kuznetsova}). On the other hand at $\beta_1^\prime<0$ and $ \beta_1^\prime<-2\tilde\beta_1$ the state 
$\vec\eta_1\propto(1,0)$, or $\vec\eta_1\propto(0,1)$ is the most profitable.

In the case of  two-band superconductivity, one can also make the model independent conclusion concerning the  order parameter form. The superconducting state can consist   only of  pairs of states belonging to same superconducting class. They are: \\
(i) $\vec\eta_1\propto(1,0)$ and $\vec\eta_2\propto(1,0)$;\\
(ii) $\vec\eta_1\propto(1,1)$ and $\vec\eta_2\propto((1,1)$;\\
(iii) $\vec\eta_1\propto(1,i)$ and $\vec\eta_2\propto(1,i)$.\\
It is easy to check that for all these pairs of superconducting states the integral over $k_x,k_y$  from the product (\ref{product}) vanishes identically.

This finding is in an exact correspondence with general statement \cite{Read} that in a translationally invariant chiral superconducting state the Hall conductivity $$\sigma_{H}({\bf q}=0,\omega )\equiv 0.$$
\subsection{Direct interband pairing}
We have to discuss now the rest of the terms in the equation (\ref{pi}) relating to the direct interband pairing  that is proportional to $\Delta_{12}$. The formal consideration of these terms making use the corresponding Landau free energy and symmetry argumentation leads to the same conclusion, namely to the absence of internal Hall effect. This case, however, the similar formal considerations are inapplicable. The point is that the interband pairing (if it exists) due to length mismatching of the oppositely directed momenta in different bands  leads inevitably to the space modulated superconducting ordering. The theory actualy cannot operate with such a space homogeneous Green functions as given by $(\ref{Green})$ and space homogeneous Landau expansions. Thus, this problem deserves some special investigation. However, one can suppose that the space ingomogeneous  order parameter distribution breaks the system translation invariance. Due to this reason the intrinsic Hall effect in chiral superconductors with interband pairing can in principle exist.

\section{Conclusion} We have studied the possibility of intrinsic Hall effect in multiband  superconductor with tetragonal symmetry 
making use the general nonconventional superconductivity theory. There was demonstrated that a superconducting state with intraband pairing including the interband pairs  transitions do not support the existence of Hall conductivity. On the other hand
this effect in principle can exist in space inhomogeneous superconducting state caused by the direct interband pairing.

\section*{Ackowledgement}This work was partly supported by the grant SINUS of the Agence Nationale de la Recherche and by European IRSES Program SIMTECH Contract No 246937.

Author is indebted to M. Zhitomirsky for the interest to this work, critical reading the manuscript and the useful discussions.


\begin{thebibliography}{20}
\bibitem{footnote} Performing this checking one needs to take into account that considering  only the equal state pairing states  we should treat the $\{\hat x,\hat y\}$ spin components of the order parameters that is $|\uparrow\uparrow\rangle$ and  $|\downarrow\downarrow\rangle$ pairing states separately from the $\hat z$ component of the order parameter,
that is $|\uparrow\downarrow+\downarrow\uparrow\rangle$ pairing state.
\end{thebibliography}
\end{document}